\documentclass[twocolumn,floats,showpacs,superscriptaddress,pre]{revtex4}
\usepackage{amsmath,amssymb,bm}
\usepackage{graphics}
\usepackage{epsfig}
\usepackage{graphicx}

\begin{document}

\title{On the Birth of the Universe and Time}
\author{Natalia Gorobey}
\affiliation{Peter the Great Saint Petersburg Polytechnic University, Polytekhnicheskaya
29, 195251, St. Petersburg, Russia}
\author{Alexander Lukyanenko}
\email{alex.lukyan@mail.ru}
\affiliation{Peter the Great Saint Petersburg Polytechnic University, Polytekhnicheskaya
29, 195251, St. Petersburg, Russia}
\author{A. V. Goltsev}
\affiliation{Ioffe Physical--Technical Institute, Polytekhnicheskaya 26, 195251, St.
Petersburg, Russia}

\begin{abstract}
A theory of the initial state of the universe is proposed within the
framework of the Euclidean quantum theory of gravity. The theory is based on
a quantum representation in which the action functional is implemented as an
operator on the space of wave functionals depending on the 4D space metric
and matter fields. The initial construction object is the eigenvalue of the
action operator in the area of the Origin of the universe with the given
values of the 3D metric and matter fields on the boundary. The wave function
of the initial state is plotted as an exponential of this eigenvalue, after
a Wick rotation in the complex plane of the radial variable of the Euclidean
4D space. An estimate of the initial radius of the universe is proposed.
\end{abstract}

\maketitle







\section{\textbf{INTRODUCTION}}

The works \cite{HH},\cite{V} laid the foundation for the development of the
idea of the quantum birth of the universe from \textquotedblleft
nothing\textquotedblright . In the work of Hartl and Hawking \cite{HH}, a
special solution of the equations of the quantum theory of gravity (the
Wheeler-De Witt equation (WDW) \cite{Wh},\cite{DeW})

\begin{equation}
\widehat{H}\psi =\widehat{H}^{i}\psi =0  \label{1}
\end{equation}%
is defined for the wave function of the universe $\psi $ in the form of a
Euclidean functional integral over all Riemannian geometries (and fields of
matter) with given boundary values on a (single) $3D$ spatial section $%
\Sigma $ (no-boundary wave function). In \cite{V}, for a similar solution, a
visual representation was proposed in the form of the amplitude of quantum
tunneling from zero to a finite radius of the $3D$ spatial section of the
universe $\Sigma $. In both cases (tunnel and no-boundary) the wave function
is calculated at the saddle point (instanton) of the Euclidean action.
Within the framework of the semiclassical approximation, in the classically
allowed region of the dynamics of the universe, one can determine the
classical time parameter. The boundary of the classically allowed region of
motion with real time is called the "bounce" point of the universe. For a
homogeneous model of the universe with a cosmological constant and a scalar
field of matter, the regularities of the formation of the inflationary stage
of the expansion of the universe immediately after the \textquotedblleft
bounce\textquotedblright\ point were studied in \cite{HHH},\cite{HHH1}. The
dependence of the position of this point on the initial value of the scalar
field at zero radius (the "south pole" of the universe) is found. The
results obtained in the semiclassical approximation for the Euclidean
functional integral can also be found directly from the WDW equation without
resorting to the functional integral \cite{HHH2}. This transformation of the
approach is caused by the fact that the Euclidean quantum theory of gravity
in terms of the functional integral \cite{GH} turns out to be untenable in
the general case (outside the semiclassical approximation) due to the sign
indefiniteness of the Euclidean action of the theory of gravity and, as a
consequence, the divergence of the integral on the space of Riemannian $4D$
metrics. Thus, time as a parameter of evolution is not defined in the modern
quantum theory of gravity (QG). In \cite{GLG}, an alternative formulation of
the QG was proposed in terms of the wave functional defined on all
pseudo-Euclidean $4D$ metrics (and matter fields) bounded by the initial and
final spatial sections $\Sigma $. In the new formulation, the wave
functional is defined by a secular equation for the action operator in the
space of world histories of the universe. Being an invariant of general
covariant transformations of the world history of the universe, it allows us
to introduce the evolution parameter as the average geodesic distance
between boundary spatial sections. In the new formulation of the QG
dynamics, the question of the initial state of the universe, which must be
determined on the initial spatial section $\Sigma _{0}$, remains open.

In this work, the quantum state of the universe on the initial spatial
section $\Sigma _{0}$ (Beginning of the universe) is found using the secular
equation for the action operator of the theory of gravity in Euclidean form.
This operator is defined on the set of $4D$ Riemannian metrics with given
boundary values for the metric and matter fields on $\Sigma _{0}$. Thus, a
unified approach is proposed for formulating the dynamics and determining
the Origin of the universe based on the operation of the Hilbert-Einstein
theory of gravity. In both cases, to determine the action operator, the
canonical form of the Hilbert-Einstein action is taken as the initial one.
However, in the case of the Euclidean action, where all four coordinates of
the $4D$ manifold are completely equal, a generalization of its canonical
structure, first formulated by De Donder and Weyl (DDW) \cite{DD},\cite{We},
is required. The generalization of the canonical quantization rules proposed
in \cite{GLG} turns out to be applicable to this generalized canonical form
of the original Hilbert-Einstein action.

In the next section, the generalized canonical form of the DDW and its
quantization are considered for a real scalar field. In the second section,
the generalized canonical form of DDW is obtained for the Hilbert-Einstein
action. On its basis, in the third section, the quantum principle of least
action (QPLA) for the Euclidean QG is formulated and the initial state of
the universe is determined.

\section{DE-DONDER-WEIL CANONICAL FORM OF THE ACTION OF A SCALAR FIELD}

As the simplest example of the generalized De-Donder-Weyl (DDV) canonical
form, consider it for the Euclidean action of a scalar field,

\begin{equation}
I_{E}\left[ \varphi \right] =\int \sqrt{g}d^{4}x\left[ \frac{1}{2}%
g^{ik}\partial _{i}\varphi \partial _{k}\varphi +V\left( \varphi \right) %
\right] ,  \label{2}
\end{equation}%
where $g_{ik}\left( x\right) $ is the Riemannian metric of the 4D manifold
with signature $(+,+,+,+)$. and $g=\det g_{ik}$. There is no distinguished
parameter of coordinate time here, and it would be natural to introduce
generalized canonical momenta for all coordinates:

\begin{equation}
p^{i}\left( x\right) \equiv \frac{\delta I_{E}}{\delta \partial _{i}\varphi
\left( x\right) }=\sqrt{g}g^{ik}\partial _{k}\varphi .  \label{3}
\end{equation}%
Using the generalized Legendre transform, we introduce the generalized
Hamilton functional,

\begin{eqnarray}
H\left[ p^{i},\varphi \right] &=&\int d^{4}xp^{i}\partial _{i}\varphi \left(
x\right) -I_{E}\left[ \varphi \right]  \notag \\
&=&\int \sqrt{g}d^{4}x\left[ \frac{1}{2g}g_{ik}p^{i}p^{k}-V\left( \varphi
\right) \right] ,  \label{4}
\end{eqnarray}%
and write action Eq. (\ref{2}) in the generalized canonical form:

\begin{equation}
I_{E}\left[ p^{i},\varphi \right] =\int d^{4}xp^{i}\partial _{i}\varphi
\left( x\right) -H\left[ p^{i},\varphi \right] .  \label{5}
\end{equation}%
It is easy to see that the extremum of action Eq. (\ref{5}) over all
variables gives the original equations for the scalar field.

The quantization of the "dynamics" of a scalar field in this generalized
canonical form is possible using the generalized canonical quantization
rules formulated in \cite{GLG}. We formulate them here, remaining in the
Euclidean form of the generalized canonical action of the scalar field Eq. (%
\ref{5}). The quantum state of the field $\varphi \left( x\right) $ is now
described by the wave functional $\Psi \left[ \varphi \left( x\right) \right]
$. For the quantum realization of generalized canonical momenta Eq. (\ref{3}%
), we take into account that the usual canonical momentum $\pi $ in
pseudo-Euclidean space-time, corresponding to the generalized coordinate $q$%
, is replaced by $-i\pi _{E}$ in the transition to the imaginary time $%
t=i\tau _{E}$, and the original action describing the dynamics in real time $%
t$ , is replaced by $-iI_{E}$. This means that the operator canonical
representation of momentum on the space of wave functions in the Euclidean
form of the theory has the form:

\begin{equation}
\widehat{\pi }_{E}=-\hbar \frac{\partial }{\partial q}.  \label{6}
\end{equation}%
In accordance with this, the generalized operator representation of the
canonical momenta of the DDW Eq. (\ref{3}) on the space of wave functionals $%
\Psi \left[ \varphi \left( x\right) \right] $ has the form \cite{GLG}:

\begin{equation}
\widehat{p}^{i}\left( x\right) \Psi =-\widetilde{\hbar }^{i}\frac{\delta
\Psi }{\delta _{i}\varphi \left( x\right) },  \label{7}
\end{equation}%
where

\begin{equation}
\widetilde{\hbar }^{i}=\hbar \epsilon ^{i},  \label{8}
\end{equation}%
and $\epsilon ^{i}$ are constant length dimensions. We will reveal the
meaning of these constant and variational derivatives in Eq. (\ref{7}) using
the lattice approximation to describe the state of the field $\varphi \left(
x\right) $. We fix coordinates in $4D$ space and introduce a set of
points $\overrightarrow{x}^{a}$ forming a lattice with a unit cell in the
form of a parallelepiped with edges $\overrightarrow{\epsilon }^{i}$ of
length $\left\vert \overrightarrow{\epsilon }^{i}\right\vert =\epsilon ^{i}$:

\begin{equation}
\overrightarrow{x}^{a}=\sum\limits_{i=1}^{4}n_{i}^{a}\overrightarrow{%
\epsilon }^{i},  \label{9}
\end{equation}%
where $a$ is the lattice node number given by a set of integers $a\equiv
\left\{ n_{i}^{a}\right\} $. Let us replace (approximate) the continuous
field $\varphi \left( x\right) $ by the set of its values $\varphi _{a}$ at
each vertex of the lattice $\overrightarrow{x}^{a}$. Let us also approximate
the wave functional $\Psi \left[ \varphi \left( x\right) \right] $ by a
function of several variables $\Psi \left( \varphi _{a}\right) $ -- the
values of the field at the lattice vertices $\varphi _{a}$. Taking into
account the connection between the variational derivative of the functional
and the partial derivative of its lattice approximation in the case of a
function of one variable $\varphi \left( t\right) $ \cite{FeHi},

\begin{equation}
\frac{\delta \Psi }{\delta \varphi \left( t_{a}\right) }=\frac{1}{%
\varepsilon }\frac{\partial \Psi }{\partial \varphi _{a}},  \label{10}
\end{equation}%
the partial variational derivative of the wave functional in Eq. (\ref{7}),
for example, in the direction $i$, is defined as follows (assuming also the
lattice approximation of the partial derivative):

\begin{equation}
\frac{\delta \Psi }{\delta _{i}\varphi \left( \overrightarrow{x}^{a}\right) }%
=\frac{1}{\epsilon _{i}}\frac{\Psi \left( \varphi _{c\neq a},\varphi \left(
\overrightarrow{x}^{a}+\overrightarrow{\epsilon }^{i}\right) \right) -\Psi
\left( \varphi _{c\neq a},\varphi _{a}\right) }{\varphi \left(
\overrightarrow{x}^{a}+\overrightarrow{\epsilon }^{i}\right) -\varphi _{a}}.
\label{11}
\end{equation}%
Then the lattice realization of the generalized canonical momentum Eq. (\ref%
{3}) will be the fraction:

\begin{equation}
\widehat{p}^{i}\left( \overrightarrow{x}_{a}\right) \Psi =-\hbar \frac{\Psi
\left( \varphi _{c\neq a},\varphi \left( \overrightarrow{x}^{a}+%
\overrightarrow{\epsilon }^{i}\right) \right) -\Psi \left( \varphi _{c\neq
a},\varphi _{a}\right) }{\varphi \left( \overrightarrow{x}^{a}+%
\overrightarrow{\epsilon }^{i}\right) -\varphi _{a}}.  \label{12}
\end{equation}%
We also write in the lattice approximation:

\begin{equation}
\partial _{k}\varphi \left( \overrightarrow{x}_{a}\right) =\frac{\varphi
\left( \overrightarrow{x}^{a}+\overrightarrow{\epsilon }^{i}\right) -\varphi
_{a}}{\epsilon ^{k}}.  \label{13}
\end{equation}%
Finally, approximating the integral in Eq. (\ref{5}) by the integral sum
over the lattice, we introduce the operator of action on the lattice
corresponding to Eq. (\ref{5}):

\begin{eqnarray}
\widehat{I}_{E}\Psi &=&-\hbar \sum\limits_{a}\prod\limits_{i}\epsilon
^{i}\sum\limits_{k}\frac{1}{\epsilon ^{k}}\left[ \Psi \left( \varphi _{c\neq
a},\varphi \left( \overrightarrow{x}^{a}+\overrightarrow{\epsilon }%
^{k}\right) \right) \right.  \notag \\
&&\left. -\Psi \left( \varphi _{c\neq a},\varphi _{a}\right) \right]
-H\left( \widehat{p}^{i},\varphi \right) \Psi .  \label{14}
\end{eqnarray}%
It is understood that the lattice approximation in all these definitions
becomes more accurate as $\epsilon ^{i}\longrightarrow 0$.

The action operator Eq. (\ref{14}), according to \cite{GLG}, allows us to
formulate the Euclidean quantum "dynamics" of a scalar field in the form of
the corresponding secular equation:

\begin{equation}
\widehat{I}_{E}\Psi =\Lambda _{E}\Psi .  \label{15}
\end{equation}%
In \cite{GLG} this formulation of dynamics is called  the quantum principle
of least action (QPLA). In what follows, we will be interested in the
eigenvalue $\Lambda _{E}$ of the action operator. We will discuss its
meaning after the formulation of a similar structure for the Riemannian
metric field $g$ and the full formulation of the QPLA for Euclidean quantum
gravity.

\section{CANONICAL DE DONDER-WEIL FORM OF THE HILBERT-EINSTEIN ACTION}

We will base the Euclidean QG on the generalized canonical form of the DDW
of the Hilbert-Einstein action, since it reflects the fact that there is no
distinguished coordinate that can be associated with coordinate time. In
this case, we are not confused by the violation of the general covariance,
which will manifest itself in the appearance of additional coordinate
conditions. We start the construction from the Euclidean form of the
Hilbert-Einstein action (at this stage we do not take into account the
matter field) \cite{GH},

\begin{equation}
I_{gE}=\frac{1}{4\pi }\int \sqrt{g}Rd^{4}x,  \label{16}
\end{equation}%
(we set $c=G=1$). Lagrangian density

\begin{eqnarray}
\pounds _{g} &=&\frac{1}{4\pi }\sqrt{g}R=\frac{1}{4\pi }\sqrt{g}g^{ik}\left(
\partial _{l}\Gamma _{ik}^{l}-\partial _{i}\Gamma _{kl}^{l}\right.  \notag \\
&&\left. +\Gamma _{ik}^{l}\Gamma _{lm}^{m}-\Gamma _{im}^{l}\Gamma
_{kl}^{m}\right) ,  \label{17}
\end{eqnarray}%
where $\Gamma _{ik}^{l}$ are Christoffel symbols \cite{MTW}. We write it as
follows:

\begin{eqnarray}
\pounds _{g} &=&\frac{1}{4\pi }\left\{ \partial _{l}\left[ \sqrt{g}\left(
g^{ik}\Gamma _{ik}^{l}-g^{il}\Gamma _{im}^{m}\right) \right] \right.  \notag
\\
&&+\sqrt{g}\left[ \left( \frac{1}{2}g^{\alpha \beta }g^{i\gamma }-g^{\alpha
i}g^{\beta \gamma }\right) \Gamma _{il}^{l}\right.  \notag \\
&&\left. +\left( \frac{1}{2}g^{\alpha \beta }g^{ik}-g^{\alpha i}g^{\beta
k}\right) \Gamma _{ik}^{\gamma }\right] \partial _{\gamma }g_{\alpha \beta }
\notag \\
&&+\left. \sqrt{g}g^{ik}\left( \Gamma _{ik}^{l}\Gamma _{lm}^{m}-\Gamma
_{im}^{l}\Gamma _{kl}^{m}\right) \right\} .  \label{18}
\end{eqnarray}%
We define the generalized momenta $P^{\gamma \left\vert \alpha \beta \right.
}$ conjugate to the components of the Riemannian metric $g_{\alpha \beta }$
as partial derivatives of the Lagrange density of action Eq. (\ref{16}) with
respect to $\partial _{\gamma }g_{\alpha \beta }$. The total divergence (the
first term in Eq. (\ref{18})) is not affected in this case, as well as the
Christoffel symbols, because the derivatives with respect to $\Gamma
_{ik}^{l}$ are reduced to the identities that define them. As a result, we
get:

\begin{eqnarray}
P^{\gamma \left\vert \alpha \beta \right. } &=&\frac{1}{4\pi }\sqrt{g}\left[
\left( \frac{1}{2}g^{\alpha \beta }g^{i\gamma }-g^{\alpha i}g^{\beta \gamma
}\right) \Gamma _{il}^{l}\right.  \notag \\
&&\left. \left( \frac{1}{2}g^{\alpha \beta }g^{i\gamma }-g^{\alpha
i}g^{\beta \gamma }\right) \Gamma _{il}^{l}\right] .  \label{19}
\end{eqnarray}%
Expression Eq. (\ref{19}) is not a tensor, so our constructions are not
covariant from the very beginning. As a consequence, we obtain additional
conditions that explicitly violate covariance. It is easy to check that the
generalized momenta Eq. (\ref{19}) obey the identities:

\begin{equation}
P^{\gamma \left\vert \alpha \beta \right. }\left( g_{\gamma \delta
}g_{\alpha \beta }-2g_{\alpha \gamma }g_{\beta \delta }\right) =0.
\label{20}
\end{equation}%
They must be taken into account when trying to solve Eq. (\ref{19}) with
respect to $\Gamma _{ik}^{l}$. We have:

\begin{eqnarray}
&&\frac{16\pi }{\sqrt{g}}P^{\gamma \left\vert \alpha \beta \right.
}g_{\gamma p}g_{\alpha q}g_{\beta r}  \notag \\
&=&\Gamma _{p\left\vert qr\right. }+\frac{1}{2}g_{qr}\left( g^{km}\Gamma
_{m\left\vert pk\right. }-g^{ik}\Gamma _{p\left\vert ik\right. }\right)
\notag \\
&&-g_{pr}g^{km}\Gamma _{m\left\vert qk\right. }.  \label{21}
\end{eqnarray}%
Then, taking into account Eq. (\ref{21}), identities Eq. (\ref{20}) lead to
additional coordinate conditions

\begin{equation}
\partial _{i}g=0.  \label{22}
\end{equation}%
This means that the canonical equality of coordinates in the DDW
representation can be achieved if the determinant of the metric tensor is
constant throughout the space. Under these additional conditions, from Eq. (%
\ref{21}) we obtain:

\begin{eqnarray}
\Gamma _{p\left\vert qr\right. } &=&\frac{16\pi }{\sqrt{g}}P^{\gamma
\left\vert \alpha \beta \right. }\left[ \left( g_{\gamma p}g_{\alpha
q}g_{\beta r}-g_{\alpha \gamma }g_{\beta p}g_{qr}\right) \right.  \notag \\
&&\left. -\frac{1}{3}g_{pr}\left( g_{\alpha q}g_{\beta \gamma }\right) %
\right] .  \label{23}
\end{eqnarray}%
We are now ready to define the generalized Hamilton functional in the DDW
representation using the generalized Legendre transform:

\begin{eqnarray}
H_{gE}\left[ P,g\right] &=&\int_{\Omega }d^{4}x\partial _{\gamma }g_{\alpha
\beta }P^{\gamma \left\vert \alpha \beta \right. }-I_{gE}  \notag \\
&=&\int_{\partial \Omega }dS_{\gamma }g_{\alpha \beta }P^{\gamma \left\vert
\alpha \beta \right. }  \notag \\
&&+\frac{1}{16\pi }\int_{\Omega }\sqrt{g}d^{4}xg^{ik}\left( \Gamma
_{ik}^{l}\Gamma _{lm}^{m}-\Gamma _{im}^{l}\Gamma _{kl}^{m}\right) .
\nonumber \\
\label{24}
\end{eqnarray}%
Taking into account Eq. (\ref{23}), the second term in the Hamilton
functional Eq. (\ref{24}) is the quadratic form of momenta:

\begin{equation}
16\pi \int_{\Omega }\frac{d^{4}x}{\sqrt{g}}\left( g_{\alpha \alpha ^{\prime
}}g_{\beta \beta ^{\prime }}g_{\gamma \gamma ^{\prime }}-g_{\alpha \gamma
}g_{\alpha ^{\prime }\gamma ^{\prime }}g_{\beta \beta ^{\prime }}\right)
P^{\gamma \left\vert \alpha \beta \right. }P^{\gamma ^{\prime }\left\vert
\alpha ^{\prime }\beta ^{\prime }\right. }.  \label{25}
\end{equation}%
After adding the matter fields to the Riemannian metric $g$, the set of
which we denote by the collective symbol $\varphi $, we can write down the
generalized canonical form of the action of the Euclidean theory of gravity
in the domain $\Omega $. We also take into account additional conditions Eq.
(\ref{20}) on the generalized canonical variables of the metric field with
the help of the corresponding Lagrange multipliers $\eta ^{\delta }$. It is
easy to see that conditions Eq. (\ref{22}) are satisfied automatically.
Finally, the action of the DDW of the Euclidean theory of gravity takes the
form:

\begin{eqnarray}
I_{E}\left[ P,g,p,\varphi ,\eta \right] &=&\int_{\Omega }d^{4}x\partial
_{\gamma }g_{\alpha \beta }P^{\gamma \left\vert \alpha \beta \right.
}+\int_{\Omega }d^{4}xp^{i}\partial _{i}\varphi \left( x\right)  \notag \\
&&-H_{gE}\left[ P,g\right] -H_{\varphi E}\left[ g,p,\varphi \right]  \notag
\\
&&\int_{\Omega }d^{4}x\eta ^{\delta }\left( g_{\gamma \delta }g_{\alpha
\beta }-2g_{\alpha \gamma }g_{\beta \delta }\right) P^{\gamma \left\vert
\alpha \beta \right. }.
\nonumber \\
\label{26}
\end{eqnarray}%
We will use this form of operation of the Euclidean theory of gravity as the
basis for the QPLA for determining the initial quantum state of the universe
in the next section.

\section{THE BEGINNING OF THE UNIVERSE AND TIME}

Let us concretize the form of the Euclidean birth region of the universe $%
\Omega $. The coordinate conditions Eq. (\ref{22}) are also satisfied in the
simplest case of a homogeneous isotropic Riemannian space, the Euclidean
space. Let us introduce in this space the spherical coordinates $x_{\alpha
}=\left( r,\theta _{A}\right) ,A=1,2,3$. The beginning of the radial
coordinate $r=0$ will be called the "south pole" of the universe in
accordance with the terminology of \cite{HHH},\cite{HHH1}. Let $\Omega $ be
a convex region covering the "south pole" and bounded by the surface $\rho
=\rho \left( \theta \right) $. The radial variable $0<r\leq \rho \left(
\theta \right) $ will be singled out as one of the spatial coordinates near
the boundary $\partial \Omega $ of the region, which we will further
\textquotedblleft join\textquotedblright\ with the time parameter of the
universe. All Riemannian metrics $g_{\alpha \beta }$ satisfying conditions
Eq. (\ref{22}) in the domain $\Omega $ will be considered related (by a
determinant-preserving bijection) to the spherical coordinates of the
Euclidean space. We introduce in these coordinates a spatial lattice $%
\overrightarrow{x}_{a}$ with constants $\epsilon ^{i}$ (see Eq. (\ref{9})).

The next step is quantization. At this step, we define the quantum version
of the action functional Eq. (\ref{26}) in the form of a difference operator
$I_{E}\left[ \widehat{P}_{a},g_{a},\widehat{p}_{a},\varphi _{a},\eta _{a}%
\right] $ on the lattice $\overrightarrow{x}_{a}$, acting in the space of
wave functionals (functions on the lattice) $\Psi \left( g_{a},\varphi
_{a},\eta _{a}\right) $ (here we agree to place the variational
differentiation operators on the right), and let us formulate the
\textquotedblleft dynamic\textquotedblright\ principle of the Euclidean
quantum theory of gravity on the domain $\Omega $ in the form of a secular
equation for the action operator:

\begin{equation}
I_{E}\left[ \widehat{P}_{a},g_{a},\widehat{p}_{a},\varphi _{a},\eta _{a}%
\right] \Psi =\Lambda _{E}\Psi .  \label{27}
\end{equation}%
This is a difference (matrix) equation for $\Psi \left( g_{a},\varphi
_{a},\eta _{a}\right) $ with given field values $\left( g_{a},\varphi
_{a},\eta _{a}\right) _{\partial
\Omega
}$ on the boundary $\partial \Omega $. Note that the first term on the right
side of Eq. (\ref{24}) in the form of a surface integral over this boundary,
when quantized, turns into a sum over lattice points satisfying the equation
$r_{a}=\rho \left( \theta _{a}\right) $, partial derivatives (finite
differences) of the wave functional $\Psi \left( g_{a},\varphi _{a},\eta
_{a}\right) $ in the radial direction (along the normal to the boundary $%
\partial \Omega $). According to the QPND formulation \cite{GLG}, the
eigenvalue $\Lambda _{E}$ of the action operator depends only on these
boundary values. We will use this eigenvalue as the basis for determining
the initial state of the universe on the boundary $\partial \Omega $, and
further fix the boundary itself by an additional extremum principle. To do
this, we recall \cite{GLG} that for nonrelativistic quantum mechanics of a
particle whose wave function is represented in exponential form

\begin{equation}
\psi \left( q,t\right) =\exp \left( \frac{i}{\hbar }R\left( q,t\right)
\right) ,  \label{28}
\end{equation}%
in the formulation of the QPND on the interval of (real) time $[0,T]$, the
eigenvalue of the action operator is equal to

\begin{equation}
\Lambda =R\left( q_{T},T\right) -R\left( q_{0},0\right) .  \label{29}
\end{equation}%
In the problem considered here, the action is Euclidean (imaginary time),
and the initial value of the wave function with phase $R\left(
q_{0},0\right) $ is absent by definition. From this it follows that the wave
function of the Beginning of the Universe should be sought in the form of an
exponential expression of the form:

\begin{equation}
\Phi _{0}\left( g_{a},\varphi _{a},\eta _{a}\right) _{\partial
\Omega
}=\exp \left[ \frac{i}{\hbar }\Lambda _{E}\left( g_{a},\varphi _{a},\eta
_{a}\right) _{\partial
\Omega
}\right] .  \label{30}
\end{equation}%
This is all we need from the spectral problem Eq. (\ref{27}). The eigenwave
functional $\Psi \left( g_{a},\varphi _{a},\eta _{a}\right) $ has the
meaning of the probability amplitude of various Euclidean $4D$ geometries
(and matter fields) inside $\Omega $ having given values on the boundary $%
\left( g_{a},\varphi _{a},\eta _{a}\right) _{\partial
\Omega
}$. But we do not need these probabilities at this stage. It is only
important that all values are finite. It can be expected that this is so for
sufficiently small values of $\rho $, where, as we know, solutions of the
classical Euclidean Einstein equations (instantons) exist. With an increase
in $\rho $, we reach the point of "return" of the classical Euclidean
solutions and exit from the "tunnel" \cite{V}. To find the spatial form of
the instanton, i.e. the function $\rho \left( \theta _{a}\right) $, we
formally \textquotedblleft go out\textquotedblright\ into a domain with real
time on the boundary $\Omega $. As real time, we will consider the
continuation of the radial variable obtained by its Wick rotation in the
complex plane $r$. Formally, this is achieved by the $3+1$ splitting of the $%
4D$ metric of Arnowitt, Deser, and Mizner \cite{ADM},

\begin{equation}
ds^{2}=\left( Ndr\right) ^{2}+g_{AB}\left( d\theta ^{A}+N^{A}dr\right)
\left( d\theta ^{B}+N^{B}dr\right) ,  \label{31}
\end{equation}%
and subsequent replacement $N\longrightarrow iN$ \cite{GLG}. It suffices to
do this on the boundary $\partial \Omega $, i.e., directly in the eigenvalue
$\Lambda _{E}$. Inside the region $\Omega $, this rotation occurs
automatically due to the constancy of the determinant $g=N^{2}\det g_{AB}$.
Taking into account that the action in the pseudo-Euclidean space $I_{L}$
with real time $t$ related to Euclidean one as follows $t=i\tau _{E}$ after
the Wick rotation is related to the Euclidean action $I_{E}$ by the relation
$I_{L}=-iI_{E}$, and, accordingly, $\Lambda _{L}=-i\Lambda _{E}$, we obtain
that the imaginary part of the exponent in Eq. (\ref{30}) (the phase of the
wave functions of the universe) is proportional to

\begin{equation}
F=-\Lambda _{E}\left( iN,g_{AB},\varphi ,\eta \right) _{\partial \Omega }.
\label{32}
\end{equation}%
The phase of the wave function in quantum mechanics is the quantum analogue
of the classical action, according to Dirac \cite{D}. To determine the
shape of the instanton corresponding to the given boundary conditions $%
\left( g_{a},\varphi _{a},\eta _{a}\right) _{\partial \Omega }$, we will
look for the minimum (extremum) of the functional Eq. (\ref{32}) with
respect to the function $\rho \left( \theta _{Aa}\right) $:

\begin{equation}
\frac{\delta F}{\delta \rho \left( \theta _{Aa}\right) }=0.  \label{33}
\end{equation}%
Let us recall in conclusion that the action Eq. (\ref{26}) and the action
operator depend on the (real) indefinite Lagrange multipliers $\eta ^{\delta
}$, which take into account the constraints Eq. (\ref{20}) that have arisen
in the DDW formalism in the theory of gravitation. We also fix them by
additional extremum conditions

\begin{equation}
\frac{\delta F}{\delta \eta _{a}^{\delta }}=0.  \label{34}
\end{equation}%
The solution of the system of equations Eqs. (\ref{33}), (\ref{34}) should
be substituted into Eq. (\ref{30}), as a result of which we obtain the wave
function the Beginning of the universe

\begin{equation}
\psi _{0}\left( g_{a},\varphi _{a},\right) _{\partial \Omega }=\exp \left[
\frac{i}{\hbar }\Lambda _{E}\left( g_{a},\varphi _{a}\right) _{\partial
\Omega
}\right] .  \label{35}
\end{equation}

\section{CONCLUSIONS}

The initial state of the universe obtained in this paper using the
generalized canonical form of the DDW in QPLA is the missing element for the
complete formulation of the dynamics of QG in terms of the wave functional
and the real time coordinate parameter \cite{GLG}. Thus, in QG, the
principle of general covariance is restored in its original sense as the
independence of the laws of quantum dynamics from an arbitrary choice of
space-time coordinates. In contrast, the QG, based on WDW, generally
excludes the use of any external coordinate parameter of time and requires
its (time) identification with one of the fundamental dynamical variables of
the theory, which inevitably destroys covariance. The noncovariance of the
initial state Eq. (\ref{35}) associated with the additional condition Eq. (%
\ref{22}) should not bother us, since the observer cannot exist in the
Euclidean region. As a solution to the system of equations Eqs. (\ref{33}), (%
\ref{34}), the function $\rho \left( \theta _{Aa}\right) $ is expressed in
terms of the boundary values of the fundamental dynamic variables $\left(
g_{a},\varphi _{a},\right) _{\partial \Omega }$. Nevertheless, we can say
that the initial $3D$ hypersurface $\Sigma _{0}$, to which we refer these
quantities, is located at a distance

\begin{equation}
r_{0}=\left\langle \frac{1}{2\pi ^{2}}\int_{\partial \Omega }\sqrt{\det
g_{AB}}d^{3}\theta \rho \left( \theta _{Aa}\right) \right\rangle _{\psi _{0}}
\label{36}
\end{equation}%
from the "south pole" of the universe in Euclidean space. Note that in
average operation, integration is carried out also over the constant
determinant $g$.This value can also be called the initial radius of the
universe.

The definition of the initial state completes the formulation of the
covariant quantum dynamics of the universe in terms of the wave functional
using arbitrary space-time coordinates \cite{GLG}. Gravitational connections
- WDW equations Eq. (\ref{1}) are not used in this formalism. However, the
new approach to the formulation of the quantum dynamics of the universe does
not preclude its description in terms of an internal parameter (multipoint,
see \cite{MTW}) of time. In this case, the question of identifying internal
time remains open. One of the options for describing the dynamics of the
universe in terms of internal time was proposed in \cite{GLG1}, where one of
the quantum numbers that arise when using the operator form of gravitational
constraints, the WDW equations, is considered as such.

\section{ACKNOWLEDGEMENTS}

We are thanks V.A. Franke for useful discussions.




\end{document}